\documentclass[conference]{IEEEtran}
\usepackage[utf8]{inputenc}
\usepackage{tikz}
\usepackage{pifont}
\newcommand{\xmark}{\ding{55}}
\usepackage{xcolor,colortbl}
\usepackage{url}
\usepackage{caption}
\usepackage{hyperref}
\newcommand{\etal}{{\em et al.}\xspace}
\usepackage{subcaption}
\usepackage{xspace}
\usepackage{adjustbox}
\usepackage{tabularx}
\newcommand{\BfPara}[1]{\vspace{1.5mm}{\noindent\bf#1.}\xspace}
\usepackage[skins]{tcolorbox}
\usepackage{xcolor}
\usepackage{makecell}
\usepackage{colortbl}
\usepackage{bbding}
\usepackage{pifont}
\usepackage{wasysym}
\usepackage{amssymb}
\usepackage{multirow}
\newcommand{\tsref}[1]{\textsection\ref{#1}\xspace}
\usepackage{tabularx}
\colorlet{lightgrey}{lightgray}

\usepackage{hyperref}
\usepackage{xcolor}
\usepackage{tcolorbox}
\usepackage{colortbl}

\tcbset{textmarker/.style={%
        enhanced,
        parbox=false,boxrule=0mm,boxsep=0mm,arc=0mm,
        outer arc=0mm,left=3mm,right=3mm,top=6pt,bottom=6pt,
        toptitle=1mm,bottomtitle=1mm}}%,oversize

\newtcolorbox{hintBox}{textmarker,
    borderline west={6pt}{0pt}{yellow},
    colback=yellow!10!white}

\newtcolorbox{blueBox}{textmarker,
    borderline west={6pt}{0pt}{blue!50!white},
    colback=blue!10!white}
    
\newtcolorbox{importantBox}{textmarker,
    borderline west={6pt}{0pt}{red},
    colback=red!10!white}
\newtcolorbox{noteBox}{textmarker,
    borderline west={6pt}{0pt}{green},
    colback=green!10!white}

\newcommand{\takeaway}[1]{\begin{blueBox} \textbf{Takeaway:} #1 \end{blueBox}}

\usepackage{multirow, booktabs}

\usepackage{cite}
\usepackage{tikz}

\IEEEoverridecommandlockouts
\usepackage{textcomp}

\makeatletter
\newcommand{\linebreakand}{%
  \end{@IEEEauthorhalign}
  \hfill\mbox{}\par
  \mbox{}\hfill\begin{@IEEEauthorhalign}
}
\makeatother

\begin{document}
    \title{Understanding the Security and Performance of the Web Presence of Hospitals: A Measurement Study}
    
    %M. kinoon, A. Alabduljabbar, R. Jang, D. Nyang, D. Mohaisen
    
%     \author{Mohammed Alkinoon}
% \affiliation{%
%   \institution{University of Central Florida}
%   \city{Orlando}
%   \state{FL}
% \email{malkinoon@Knights.ucf.edu}

%     \author{Abdulrahman Alabduljabbar}
% \affiliation{%
%   \institution{University of Central Florida}
%   \city{Orlando}
%   \state{FL}
% \email{jabbar@Knights.ucf.edu}

% \author{Rhongho Jang}
% \affiliation{%
%   \institution{Wayne State University}
%   \city{Detroit}
%   \state{MI}
%   \country{United States}}
% \email{r.jang@wayne.edu}

% \author{DaeHun Nyang}
% \affiliation{%
%   \institution{Ewha Womans University}
%   \city{Seoul}
%   \country{Republic of Korea}}
% \email{nyang@ewha.ac.kr}

% \author{David Mohaisen}
% \affiliation{%
%   \institution{University of Central Florida}
%   \city{Orlando}
%   \state{FL}
% \email{mohaisen@ucf.edu}

\author{
\IEEEauthorblockN{Mohammed Alkinoon}
\IEEEauthorblockA{\textit{Department of Computer Science} \\
\textit{University of Central Florida}\\
Orlando, FL, USA \\
malkinoon@Knights.ucf.edu}
\and
\IEEEauthorblockN{Abdulrahman Alabduljabbar}
\IEEEauthorblockA{\textit{Department of Computer Science} \\
\textit{University of Central Florida}\\
Orlando, FL, USA \\
jabbar@Knights.ucf.edu}
\and
\IEEEauthorblockN{Hattan Althebeiti}
\IEEEauthorblockA{\textit{Department of Computer Science} \\
\textit{University of Central Florida}\\
Orlando, FL, USA \\
hattan@Knights.ucf.edu}
\linebreakand
\IEEEauthorblockN{Rhongho Jang}
\IEEEauthorblockA{\textit{Department of Computer Science} \\
\textit{Wayne State University}\\
Detroit, MI, USA \\
r.jang@wayne.edu}
\and
\IEEEauthorblockN{DaeHun Nyang}
\IEEEauthorblockA{\textit{Department of Computer Science} \\
\textit{Ewha Womans University}\\
Seoul, Republic of Korea \\
nyang@ewha.ac.kr}
\and
\IEEEauthorblockN{David Mohaisen}
\IEEEauthorblockA{\textit{Department of Computer Science} \\
\textit{University of Central Florida}\\
Orlando, FL, USA \\
mohaisen@ucf.edu}
}

\maketitle

\begin{abstract}
The recent transformation of healthcare medical records from paper-based to digital and connected systems raises concerns regarding patients' security and online privacy. For instance, sensitive personal information, such as patients' names, addresses, and social security numbers, may be targeted due to the lack of proper security and privacy mechanisms. 

Using a total of 4,774 hospitals categorized as government, non-profit, and proprietary hospitals, this study provides the first measurement-based analysis of hospitals' websites and connects the findings with data breaches through a correlation analysis. We study the security attributes of three categories, collectively and in contrast, against domain name-, content-, and SSL certificate-level features. We find that each type of hospitals has a distinctive characteristic of its utilization of domain name registrars, top-level domain distribution, and domain creation distribution, as well as content type and HTTP request features. Security-wise, and consistent with the general population of websites, only 1\% of government hospitals utilized DNSSEC, in contrast to 6\% of the proprietary hospitals. Alarmingly, we found that 25\% of the hospitals used plain HTTP, in contrast to 20\% in the general web population. Alarmingly too, we found that 8\%-84\% of the hospitals, depending on their type, had some malicious contents, which are mostly attributed to the lack of maintenance. 

We conclude with a correlation analysis against 414 confirmed and manually vetted hospitals' data breaches. Among other interesting findings, our study highlights that the security attributes highlighted in our analysis of hospital websites are forming a very strong indicator of their likelihood of being breached.  Our analyses are the first step towards understanding patient online privacy, highlighting the lack of basic security in many hospitals'  websites and opening various potential research directions.

\end{abstract}

% \ccsdesc[500]{Information systems}
% \ccsdesc[300]{World Wide Web; Web mining}
% \ccsdesc{Healthcare Security and Privacy}
% \ccsdesc[100]{Usability in security and privacy}
%\settopmatter{printacmref=false}


\section{Introduction}
Electronic Health Records (EHR) are longitudinal electronic patient health information records, which include patient demographics, progress notes, health problems, medications, vital signs, medical history, immunizations, laboratory data, etc.~\cite{BenefitsOfEHR2}. 
The adoption of EHRs has led to improved accessibility of healthcare information for patients and providers, resulting in higher quality patient care and more efficient coordination between hospitals. However, despite these benefits, the transformation to EHRs has also raised privacy and security concerns, particularly when EHR data is retrievable through website systems. EHRs centralize sensitive patient data, which can make them a prime target for cybercriminals seeking to steal or exploit this information. Additionally, EHRs can be accessed and shared across multiple healthcare providers, increasing the risk of data breaches and unauthorized access. Moreover, the implementation of EHR systems can introduce new vulnerabilities that may be exploited by cybercriminals.

For instance, vulnerabilities and software exploits in the healthcare domain have become a central focus of targeted cyber attacks \cite{kamoun2014human}, which can result in devastating data breaches. Given the sensitivity of private healthcare data, the unauthorized and illegitimate disclosure of this information can have catastrophic consequences \cite{InsightsandImplications,AlkinoonCM21}.

% For instance, the increasing vulnerabilities and software exploitations in the healthcare domain are central in targeted cyber-attacks \cite{kamoun2014human}. 
% Exploiting such vulnerabilities in this domain may lead to ``data breaches''. Given the sensitivity of such private healthcare data, the ramifications of this unauthorized and illegitimate disclosure are catastrophic~\cite{seh2020healthcare,AlkinoonCM21}. 

Understanding the effect of healthcare data breaches is essential, and efforts in the literature classified those breaches into internal and external breaches. Internal breaches are commonly caused by human errors, particularly among healthcare employees. In contrast, external breaches, which are the more critical type, are caused by an unauthorized third party involved in the theft of private health records through hacking of the web-based user and healthcare provider-facing systems~\cite{wikina2014caused}. Cybercriminals typically commit these incidents, making their effect an open question, with no accurate assessment of their cost. For instance, adversaries involved in external breaches may aim to steal sensitive records and demand a ransom or sell those records for hundreds of dollars per single patient on the dark web~\cite{smith2016examining,walker2018systematic}.

Given the importance of understanding data breaches in healthcare and the role of web technologies in enabling a significant part of the attack surface, this study is dedicated to analyzing the commonalities and differences among three types of hospitals: government public hospitals, non-profit hospitals, and proprietary hospitals. Namely, we analyze the websites and patients' portals for security configurations and common privacy practices. 
We note that compromising patients' portals allows the attacker to obtain sensitive information regarding the patient's records, including diagnoses, treatment records, hospital visits, and future appointments, alongside personal information. To the best of our knowledge, this work is the first in this direction, associating actual hospital potential exploitations and data breaches with website security and privacy configurations.

Our analysis is based on a total of 4,774 hospital websites grouped into three major hospital categories: government public hospitals, non-profit hospitals, and proprietary (private) hospitals. For our measurement assessments, we conduct both domain-level and content-level analyses to understand the similarities and differences among website attributes. 

Our analysis is multi-faceted and covers a range of features by examining and comparing the website's domain SSL certificates, creation date, HTTP requests, page size, content type, average load time, and malicious activity association. The features explored in this analysis are particularly lightweight and do not require deep analyses of contents but rather focus on meta-attributes, making our analysis techniques more generalizable to large-scale measurements. We further investigated the security attributes of these websites by exploring their association with malicious behaviors, including an assessment of the domain-based and content-based malicious behaviors of those websites and associated trends and characteristics. 

To understand the implications of those characteristics, we further study their correlation with a manually vetted dataset of recently disclosed data breaches provided by the U.S. Department of Health and Human Services, the Office for Civil Rights (OCR)~\cite{OCR}. Leveraging information regarding the websites and associated breaches, we extracted the commonalities among hospitals' websites targeted with those data breach attacks towards their modeling and characterization. We believe that this work is the first step towards understanding website attributes that may lead to breaches and enable future research on vulnerability prediction and detection.

\BfPara{Research Questions} We aim to answer an overarching single question: {\bf Is there any difference between the different categories of hospital websites with respect to the studied features across content, performance, and security?} We break the question down into the following quantifiable questions. 
\begin{itemize}

\item {\bf RQ1.} How different are different hospitals with their use of domain, content, and transport layer features? We answer this question by comparatively exploring the domain-level features (section~\ref{sec:domain}), including the domain name registrar, top-level domain distribution, domain creation distribution, and content-level features, including the content type and HTTP request features (section~\ref{sec:content}). 
\item {\bf RQ2.} What are the main security characteristics of hospital websites, and how do they differ across types? We answer this research question by exploring the DNSSEC deployment of hospitals and its contrast (section~\ref{sec:dnssec}, SSL certificate features and properties (section~\ref{sec:sslcetrificate}), maliciousness characteristics against various engines (section~\ref{sec:malicious}), and data breaches association (section~\ref{sec:databreach}).
\end{itemize}

\BfPara{Contributions and Findings} Given the lack of any systematic work on understanding the characteristics of hospitals' presence on the web and their associated security and performance attributes, this study sets out to explore these hospitals' web presence across a range of attributes. Moreover, through a comparative analysis, this work uncovers the differences and similarities between the Government, Non-profit, and Proprietary hospitals in the United States. Our analysis is conducted across three dimensions: security, contents, and domains. To this end, our contributions are as follows:
% Our contributions are as follows.

\begin{enumerate}
    \item \BfPara{Domain-level Analysis (\tsref{sec:domain})} Domain names are the gateway to websites, and they are essential to understanding various coarse-grained and easy-to-obtain features of those domains and entities behind them. To this end, we conduct a domain name registrar and top-level domain analysis to uncover websites/hospitals' characteristics and to contrast them. We uncover the affinities in the choice between websites and registrars, top-level domain choice, and domain creation dates. Among other interesting findings, we observe that the number of websites for government and non-profit hospitals has been declining in recent years, hinting at the aggressive proprietary healthcare system.
  
  \item \BfPara{Content-level Analysis (\tsref{sec:content})} We examined the contents of hospitals' websites for a deeper look into their utilized content types, size, and employed security features. Through this analysis, we found the gap in employing different content types, such as images and scripts, in those websites, which affects the various performance metrics, including loading times. More interestingly, and rather surprisingly, we found that 6\% of proprietary hospitals use the Domain Name System Security Extension (DNSSEC), in comparison to less than 1\% of the government and non-profit hospitals. 
  
 \item \BfPara{SSL Certificate-level Analysis (\tsref{sec:sslcetrificate})} Certificates are essential for website authentication and to facilitate web content encryption at the transport layer, providing a secure application medium. We investigate the HTTPS protocol configurations, associated SSL features, and the SSL certificate validity of hospitals' websites. We categorize websites based on certificate authority affinity, utilized algorithms, and certificate validity. Among other interesting findings, our investigation uncovers that more than 25.25\% of hospital websites are still using the insecure HTTP protocol. Further, among websites that utilize HTTPS, up to 23\% of the SSL certificates are invalid.
  
  \item \BfPara{Malicious Activities Analysis (\tsref{sec:malicious})} Because of their complex nature, unintended weaknesses might emerge due to the agglomeration of third-party code and the utilization of various shared pieces of infrastructure in hospital websites. To understand this dimension, we utilize various scanning tools to explore those websites' malicious activities at the domain and content levels. Among other interesting findings, we uncover that a large portion of websites contains malicious content and are associated with malicious behaviors.
  
\item \BfPara{Data Breaches Analysis (\tsref{sec:databreach})} Data breaches are inevitable. But what (in the correlation sense) makes a website prone to data beach? We explore this question by correlating and associating the hospitals to recently reported data breach incidents and uncover that non-profit hospitals are more likely to be involved in data breach incidents. We demonstrate the most important attributes contributing to data breach incidents, including hosting malicious codes, a large number of images, etc. 
\end{enumerate}

\section{Related Work}
Limited work is presented on analyzing the security configurations of healthcare providers. To contextualize our work, we review previous works on website content, security analysis, and healthcare data breach analysis.

\BfPara{Websites Analysis} Over the past few years, there has been a drastic increase in the development and utilization of online services and web applications. Paralleled with this rise has been an increasing concern over the privacy and security of these online services and applications, e.g., different components can be compromised, putting their users at risk. Chung \etal~ \cite{TaejoongChung} offered the first in-depth analysis of incorrect certifications in the online Public Key Infrastructure (PKI), showing that most PKI certificates are invalid. The same study scrutinized the origin of the invalid SSL certificates and summarized that the preponderance of the invalid certificates was generated by end-user devices, with a periodical renewal of new self-signed certificates. SSL certificates have been investigated for website risk and vulnerability analysis \cite{DoowonKim, DoowonKim2, DoowonKim3, LiangZhang, MishariAlMishari, BumJunKwon,JeremyClark,TaejoongChung,Yuting, BerkowskyH17,BatesPNHTBA, AlrawiM16, alabduljabbar2022privacy}. For instance, Meyer \etal~\cite{UlrikeMeyer} examined the SSL certificates' content and information to distinguish between phishing and benign websites. Alabduljabbar \etal~\cite{alabduljabbar2022certificates, alabduljabbar2022nfl} explored the SSL certificate-based structural differences between free and premium content websites and highlighted that 35.85\% of the free websites' certificates have significant security issues, with 17\% invalid, 7\% expired, and 12\% with mismatched domain names.

\BfPara{Data Breaches Analysis} 
Several studies have recently investigated data breaches in the healthcare industry \cite{AlkinoonCM21, AlkinoonOMM21,Multi-Campus,InsightsandImplications}. For instance, Seh \etal~\cite{InsightsandImplications} conducted a comprehensive analysis of HIPPA data breach reports. Their study highlights that hacking incidents, unauthorized access (internal), theft or loss, and improper disposal of unnecessary data are the main disclosure types of protected healthcare information. Moreover, the authors applied the Simple Moving Average (SMA) and Simple Exponential Smoothing (SES) time series methods on the data to determine the trend of healthcare data breaches and their cost to the healthcare industry. Choi \etal~\cite{Ads} estimated the link between data breaches and hospital advertising spending, studying the period of the two years following the breach and finding hospitals had much higher advertising expenditures. Siddartha \etal~\cite{MaskingTechniques} found that the healthcare industry is being targeted for two main reasons: being a rich source of valuable data and its weak defenses. 

Siddartha and Ravikumar~\cite{MaskingTechniques} suggested that the security techniques employed in the healthcare industry miss data analysis improvements, e.g., data format preservation, data size preservation, and other factors. Luis \etal~\cite{Multi-Campus} defined DNS queries and TLS/SSL connections to identify the dangers encountered inside a hospital environment without disrupting the functioning network using two years of collected data. Another line of work, the 2022 Data Breach Investigations Report (DBIR)~\cite{DBIR2022}, investigates healthcare breaches among other industries. Based on the report, healthcare suffered 849 incidents, with 571 confirmed data disclosure in 2022. The report summarized various findings and determined that external actors are behind 61\% of data breaches while 39\% of data breaches involved internal actors. Furthermore, according to the same report, financial gain is the highest motive for attackers 95\%, followed by espionage 4\%. 

\BfPara{This Work} We focus on the online presence of healthcare organizations, diving into the security configurations and features of the hospitals' websites and patient portals through contrast across three major hospital types and the correlation of such features with reported data breaches.

% \subsection{Web content}

\section{Dataset, Pipeline and Research Questions}
For this study, we utilized an authentic dataset of U.S. hospitals obtained from the Homeland Infrastructure Foundation-Level Data (HIFLD) \cite{dataSource}. 
The dataset contains hospitals distributed among the 50 U.S. states, Washington D.C., and U.S. territories of Puerto Rico, Guam, American Samoa, Northern Mariana Islands, Palau, and the Virgin Islands.

We categorized the hospitals in the dataset into three categories: government, non-profit, and proprietary. The government hospitals include federal, district, local, and state hospitals. Non-profit hospitals are those operated using charities according to the Internal Revenue Service (IRS)~\cite{NonprofitInfo}. The proprietary hospitals are those owned and operated for profit by individuals, partnerships, or, in most cases, corporations \cite{ProprietaryInfo}. 
Overall, we had 1,034 governmental hospitals, 2,187 non-profit hospitals, and 1,550 proprietary hospitals. 

\begin{figure}[t]
    \centering
        \includegraphics[width=0.47\textwidth]{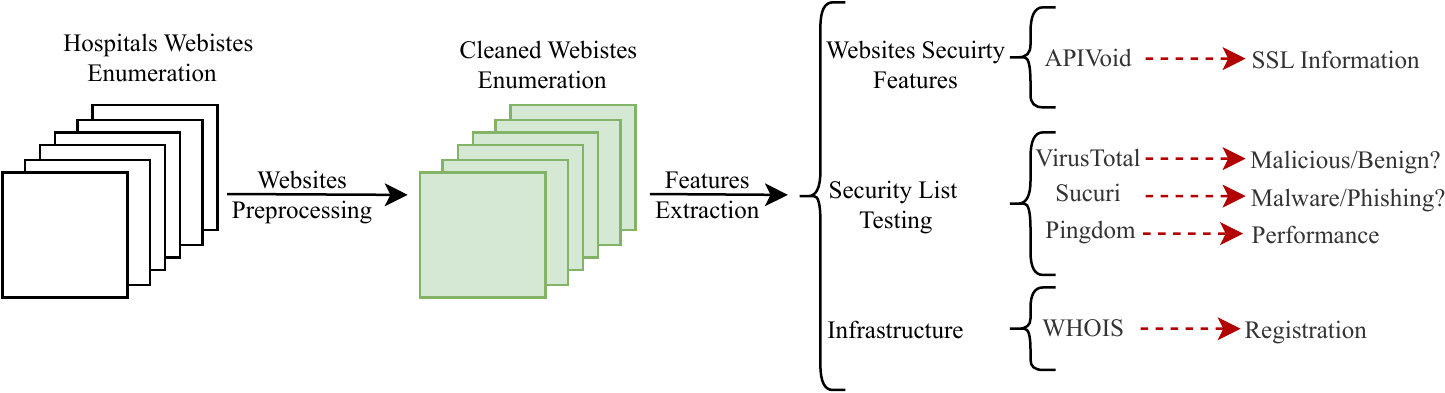}
        \caption{Our pipeline with the steps taken in websites crawling and data augmentation against various dimensions: SSL, maliciousness, vulnerability, performance, and domain attributes.}
        \label{fig:DataPipeline}
\end{figure}

\BfPara{Websites Preprocessing and Crawling} Our study involved the use of website crawling to systematically gather information from websites and incorporate it for our further analysis. Figure~\ref{fig:DataPipeline} illustrates the overall process we followed, starting with website enumeration to identify the websites we needed to crawl. Next, we conducted website preprocessing, which involved removing irrelevant websites (i.e., websites with irrelevant contents to the scope of the study) and non-functioning websites. Finally, we performed feature extraction to extract attributes from the websites that we used for our further analysis.

To conduct our analysis, we introduced a data augmentation step. Upon crawling the websites associated with each hospital, we enriched the collected data with additional attributes, including the following.
\begin{itemize}
    \item {\bf SSL Attributes.} To extract SSL certificate information such as mismatched domains, SSL expiration dates, and certificate validity, we utilized APIVoid~\cite{apivoid}, a framework that offers cyber threat analysis and detection capabilities.
    
    \item {\bf Maliciousness Attributes.} To analyze the maliciousness of hospital online content, we employed VirusTotal API~\cite{VirusTotal}, an online service that aggregates results from over 70 scanning engines.
    
    \item {\bf Vulnerability Attributes.} To examine each website for vulnerabilities and identify any malicious code, we utilized Sucuri~\cite{Sucuri}, a service that tests websites against multiple known malware and blacklisting lists.
    
    \item {\bf Performance Attributes.} To evaluate website performance and availability, we utilized Pingdom \cite{Pingdom}, a global monitoring software for websites.
    
    \item {\bf Domain Attributes.} To determine ownership and DNSSEC information for each website, we utilized WHOIS \cite{whois}, an Internet resource ownership database, and queried each website's creation date.
\end{itemize}
%\cced{1}  \cced{2}  \cced{3}  \cced{4}  
Overall, the steps of websites crawling and data augmentation allowed us to extract two types of information: \ding{172} website content data such as images, fonts, HTML, CSS, scripts, XHR, and redirects, and \ding{173} performance metrics such as page size, load time, and the number of requests. %\cced{5} 
   
% \warning{Put a pipeline and reference in the website crawling part.}

%\subsection{Research Questions}

%\warning{Fill in the research questions here.}

% \item {\bf RQ3.} What is the prevalence of DNSSEC deployment in nonprofit, government, and proprietary hospitals, and how does it differ across these different types of organizations?
% We answer this question by... 

% \item {\bf RQ4.} What is the prevalence of SSL certificate security issues, such as mismatched, invalid, or expired certificates, in nonprofit, government, and proprietary hospitals? We answer this question directly by analyzing the SSL certificate features comparatively in section~\ref{sec:sslcetrificate}. 

% \item[RQ5.] What are the differences in SSL certificate security measures among government, proprietary, and non-profit hospitals' websites, and how might these differences impact patients' trust and confidence in the hospitals' online presence?
% \item[RQ6.] How do government, proprietary, and non-profit hospitals' websites compare in terms of their susceptibility to malware and phishing attacks
% \item[RQ7.] What are the differences in data breach surfaces among government, non-profit, and proprietary hospitals?
% \item[RQ8.] What are the differences in types of data breaches experienced by government, non-profit, and proprietary hospitals?
% \item[RQ9.] What is the relative importance of domain- and content-level attributes in distinguishing hospital websites associated with data breaches

\section{Websites Analysis}

% \warning{Define the analysis metrics and dimensions beforehand.}
% To conduct our analysis, we used the following metrics and dimensions.

% \BfPara{Domain-level analysis} Domain-level analysis refers to the process of examining a domain name or URL to gain insights into the characteristics, properties, and potential risks associated with a particular website or online resource. Below are the metrics and dimensions used in our analysis.

% \begin{itemize}
% \item Domain Name Registrar 
% \item Top Level Domain (TLD)
% \item Domain Creation
% \end{itemize}

To understand the online presence and structural differences between hospitals of different categories, we conducted a range of analyses of their websites:  domain-, content-, SSL certificate-, and malicious activities-based analyses. In the following, we review the findings from those analyses.

\begin{table}[t]
\centering
\caption{The hospitals' website URLs correspond to domain registrar organizations. Notice that \textit{Network Solutions} and \textit{GoDaddy} are the most prominent in the list, with up to 67.65\% associated URLs.}
\label{Tab:domain_registrar}
\begin{adjustbox}{width=\columnwidth,center}
\begin{tabular}[t]{lrrrrrr}
\Xhline{2\arrayrulewidth}
\multirow{2}{*}{Domain Registrar} & \multicolumn{2}{c}{Government} & \multicolumn{2}{c}{Non-profit} & \multicolumn{2}{c}{Proprietary} \\
\cline{2-7}
  & \# & \% & \# & \% & \# & \%   \\
\Xhline{2\arrayrulewidth}
Network Solutions\_LLC   & 330 & 35.22     & 956  & 44.90    & 345 & 22.91 \\
GoDaddy.com\_LLC         & 314 & 33.51     & 621 & 29.17     & 527 & 34.99 \\
MarkMonitor\_Inc       & 1  & 0.11       & 19  & 0.89      & 156 & 10.36 \\
eNom\_LLC                & 28 & 2.99       & 51 & 2.40       & 159 & 10.56\\
Register.com\_Inc       & 24  & 2.56      & 43  & 2.02      & 14  & 0.93 \\
NAMECHEAP\_Inc            & 13  & 1.39      & 37  & 1.74      & 24 & 1.59  \\
CSC CORPORATE\_Inc   & 2 & 0.21     & 51  & 2.40    & 19  & 1.26 \\
Tucows\_Inc          & 37  & 3.95    & 26 & 1.22    & 10  & 0.66 \\
Other  & 188 & 20.06                       & 325 & 15.27       & 252  & 16.73 \\
\Xhline{2\arrayrulewidth}
\end{tabular}
\end{adjustbox}
\end{table}
%%%%%%%%%%%%%%%%%%%%%%%%%%%%%%%%%%%%%%%%%%%%%%%%%%%%%%%%%%%%%%%%%%%%%%%%%%%%%%%%%%%%%%%%%%%

\begin{table*}[t]
\centering
\caption{Top-Level Domain comparison between the Government, Non-profit, and Proprietary hospitals.}
\label{Tab:tld}
\begin{tabular}{lrrrrrrr}
\Xhline{2\arrayrulewidth}
       Type     & .org     & .com     & .gov    & .net    & .mil    & .edu    & .us      \\
\Xhline{2\arrayrulewidth}
Government  & \cellcolor{green!20}48.45\% & \cellcolor{green!20}36.56\% & \cellcolor{green!10}4.35\% & \cellcolor{green!10}4.35\% & \cellcolor{green!5}2.90\% & \cellcolor{green!5}2.42\% & \cellcolor{green!5}0.48\%  \\
Non-profit  & \cellcolor{green!40}67.49\% & \cellcolor{green!15}28.81\% & \cellcolor{green!2}0.05\% & \cellcolor{green!5} 1.92\% & \cellcolor{green!0}0.00\% & \cellcolor{green!3}1.37\% & \cellcolor{green!3} 0.27\%  \\
Proprietary & \cellcolor{green!10}7.81\%  & \cellcolor{green!40}87.35\% & \cellcolor{green!0}0.00\% & \cellcolor{green!6}3.48\% & \cellcolor{green!0}0.00\% & \cellcolor{green!2}0.13\% & \cellcolor{green!5}0.45\% \\
\Xhline{2\arrayrulewidth}
\end{tabular}
\end{table*}
%%%%%%%%%%%%%%%%%%%%%%%%%%%%%%%%%%%%%%%%%%%%%%%%%%%%%%%%%%%%%%%%%%%%%%%%%%%%%%%%%%%%

\subsection{Domain-level Analyses}\label{sec:domain}
% Domain names are essential properties for any organization, as they are used for branding, allowing them both presence and search engine optimization (SEO). Thus, we start the website analysis using the domain name information: domain name registrar, top-level domain, and domain creation date.

The domain name is a crucial asset for any organization, serving as a key element in their branding efforts and providing them with a strong online presence and Search Engine Optimization (SEO) benefits. Therefore, in order to kickstart our website analysis, we begin by examining the domain name details, including the domain name registrar, the top-level domain, and the domain creation date.

\BfPara{Domain Name Registrar} The domain name registrar is an organization that manages the reservation of Internet domain names, as well as the assignment of IP addresses for those domain names~ \cite{domainInfo}, and certain registrars tend to be more lax with their security provisions and policies~\cite{coull2012understanding, cova2010analysis, van2008economics}. Analyzing the domain name registrar is crucial in evaluating a website's overall security and reliability. This is because the registrar provides important information about the organization's online presence and security measures. The level of security provisions and policies of the registrar can vary, which can impact the website's security and trustworthiness. Examining the domain name registrar can provide valuable insights into the organization's security approach and help assess potential risks associated with the domain name. In addition, understanding the registrar can shed light on the organization's online strategy and web hosting arrangements, which can further inform the analysis of the website's structure and performance. Table~\ref{Tab:domain_registrar} shows the breakdown of domain registrar organizations by hospital type. Notably, \textit{Network Solutions} and \textit{GoDaddy} are the most prominent registrars, accounting for up to 67.65\% of the domains. Additionally, we observe that although \textit{Mark Monitor} and \textit{eNom LLC} are relatively absent from government and non-profit websites, they contribute to 20.92\% of proprietary hospital websites. 

% Table~\ref{Tab:domain_registrar} shows the corresponding domain registrar organizations broken down per hospital type. We notice that \textit{Network Solutions} and \textit{GoDaddy} are the most prominent registrars, with up to 67.65\% of the domains. Moreover, we observe that \textit{Mark Monitor} and \textit{eNom LLC} contribute to 20.92\% of proprietary hospital websites, despite being relatively absent from the government and non-profit websites.

\BfPara{Top Level Domain}
The Top-Level Domain (TLD) is the ``extension'' of a domain name. Besides branding, TLD plays an essential role in the Domain Name System (DNS) lookup and helps classify and communicate the purpose of domain names. Examples of TLDs include ".com," ".org," ".net," and ".edu." The TLD provides information about the website's purpose, organization type, or geographic location. For instance, ".com" is commonly used for commercial websites, ".org" for non-profit organizations, ".edu" for educational institutions, and ".gov" for government websites. Understanding the TLD can provide insights into the website's intended audience and the type of content or services it provides. In addition, analyzing the TLD can help identify potential security risks associated with the website. For example, certain country-specific TLDs are known to be associated with malicious activities, and websites using such TLDs may be more likely to pose security threats. Moreover, some websites may use TLDs that are misspelled or similar to well-known TLDs in an attempt to deceive visitors and carry out fraudulent activities. By analyzing the TLD, we can identify such risks and take appropriate measures to mitigate them. Therefore, analyzing the TLD is a crucial step in evaluating a website's overall security and reliability. The Internet Corporation for Assigned Names and Numbers (ICANN) is responsible for the authority over all TLDs on the internet and delegates these TLDs' responsibility to various organizations\cite{tldInfo} Table~\ref{Tab:tld} shows the TLD comparison between the hospital categories in our dataset. We observe that  \textit{``.org''} is the most dominant TLD for the government (48.45\%) and non-profit (67.49\%), while it is relatively absent in the proprietary hospitals (only 7.81\%). On the other hand, \textit{``.com''} is dominant for proprietary hospitals (87.35\%) compared to (36.56\% and 28.81\%) in government and non-profit hospitals, respectively. 
We also notice that 92.15\% of the hospitals' websites, in the aggregate, have \textit{``.com''} or \textit{``.org''}. 
Despite our common beliefs, we surprisingly uncover that only 4.35\% of the government websites use the \textit{``.gov''} TLD.

\begin{figure}[t]
    \centering
        \includegraphics[width=0.47\textwidth]{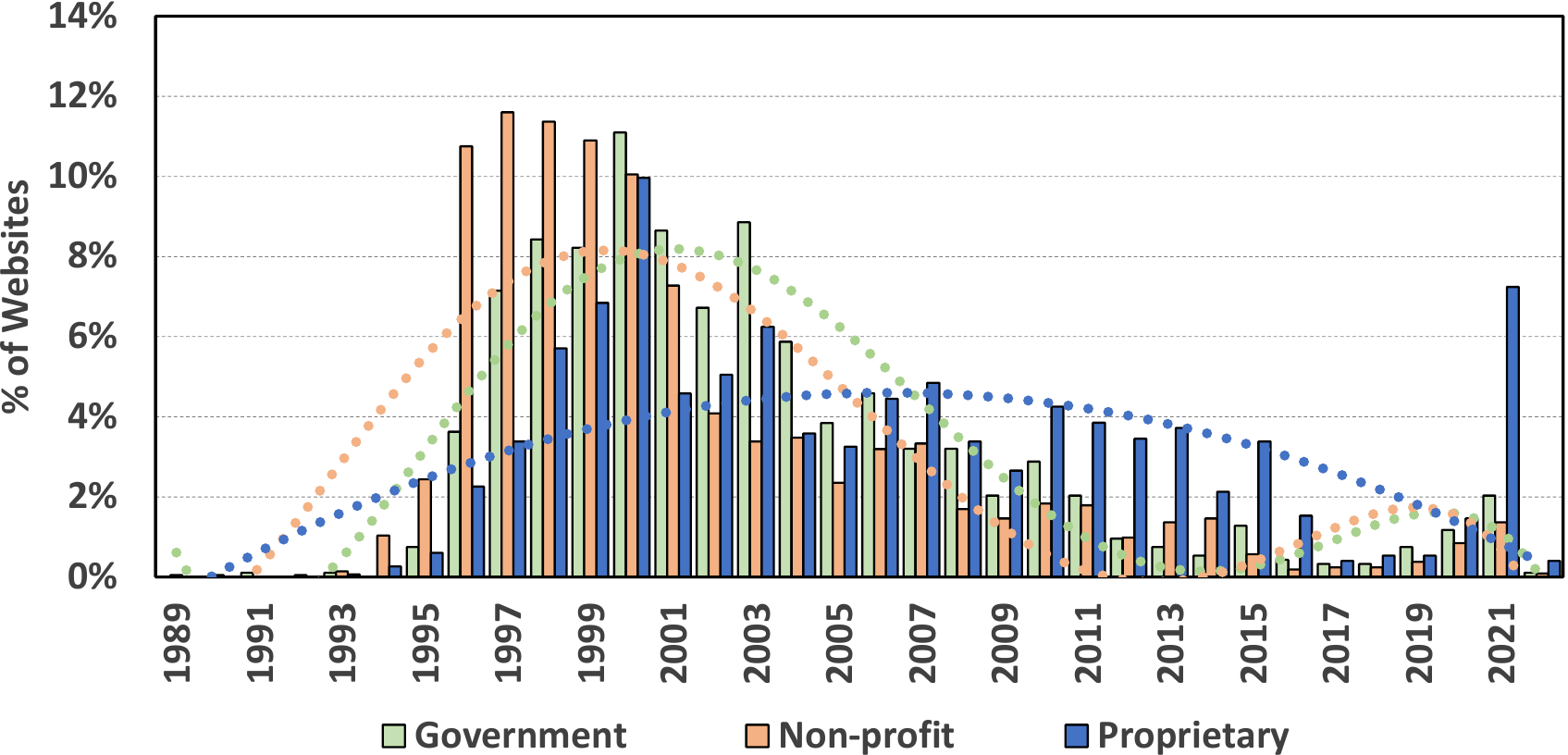}
        \caption{The domain creation date temporal analysis between the three hospital categories. Dot lines are moving averages.}
        \label{fig:creationdate}

\end{figure}

\BfPara{Domain Creation}
The domain creation date refers to the date on which a specific domain name was initially registered with a domain name registrar. It is a crucial piece of information because it can provide insights into the website's history and online presence. A website that has been registered for a longer duration may be more established and have a more significant online presence than a newer website. Analyzing the domain creation date can be useful in identifying potentially fraudulent or malicious websites. For instance, a website that has been recently registered may be more likely to be a part of a phishing scam or a fraudulent scheme. Therefore, analyzing the domain creation date can be a valuable step in evaluating the website's overall credibility and potential security risks. Figure~\ref{fig:creationdate} shows the domain creation date of hospitals in different categories. As shown, both government and non-profit hospitals' websites emerged in a similar period (1995 -- 2009), with a declining trend after 2009. However, the emergence of proprietary hospital websites is steady, with a rapid increase in their numbers in 2021. 

\begin{table*}[t]
\centering
\caption{Content-type comparison between the Government, Non-profit, and Proprietary hospitals.}\label{tab:content_type}
\scalebox{0.99}{
\begin{tabular}{lrrrrrrr}
\hline
Category    & CSS    & Font   & HTML   & Image   & Redirect & Script  & XHR    \\
\hline
Government & \cellcolor{green!10} 6.05\% & \cellcolor{green!13}8.18\% & \cellcolor{green!5}2.71\% & \cellcolor{green!40}40.55\% & \cellcolor{green!20}17.37\% & \cellcolor{green!20}20.34\% & \cellcolor{green!10}4.79\% \\

Non-profit & \cellcolor{green!8}5.30\% & \cellcolor{green!11}7.80\% & \cellcolor{green!5}2.86\% & \cellcolor{green!35}38.01\% & \cellcolor{green!18}16.11\% & \cellcolor{green!25}27.01\% & \cellcolor{green!7}2.91\% \\

Proprietary & \cellcolor{green!10}6.63\% & \cellcolor{green!15}9.61\% & \cellcolor{green!8}3.53\% & \cellcolor{green!25}28.05\% & \cellcolor{green!12}13.76\% & \cellcolor{green!35}33.12\% & \cellcolor{green!15}5.29\%\\
\hline
\end{tabular}}
\end{table*}

%%%%%%%%%%%%%%%%%%%%%%%%%%%%%%%%%%%%%%%%%%%%%%%%%%%%%%%%%%%%%%%%%%%%%%%%%%%%%%%%%%%%

%\noindent\textit{\textbf{\underline{Key Takeaway:}}}

\takeaway{({\bf RQ1.}) While the number of websites for government and non-profit hospitals has been declining in recent years, proprietary hospitals have been growing significantly, particularly in 2021. Moreover, despite being government-supported, most government hospitals do not have \textit{``.org''} top-level domain.}

\subsection{Content-level Analyses}\label{sec:content}
To analyze the content differences between the different categories of hospitals, we crawled the hospitals' websites using Pingdom \cite{Pingdom}, obtaining the HTTP request information and all associated files; scripts, images, HTML, and CSS files.

\BfPara{Content Type}
On the structural level, Table~\ref{tab:content_type} shows the distribution of the file type among the three hospital categories. XHR is an API used as an object to interact with servers and exchange data between servers and web browsers. Containing \textit{``XHR''} is prominent among the government and proprietary hospital websites, with 10.08\% combined. Overall, the file type distribution is similar for all categories, except for \textit{``Image''} and \textit{``Script''}. The \textit{``Script''} content, which is defined as a computer program for adding dynamic capabilities to a website, is used most among the proprietary hospitals with 33.12\%. The \textit{``Redirect''} content, on the other hand, which is a website feature that sends a user from the current URL to another server, is applied more in the government (17.37\%) and non-profit (16.11\%) and the least in the proprietary category (13.76\%).

\begin{figure}[t]
    \centering
    \includegraphics[width=0.45\textwidth]{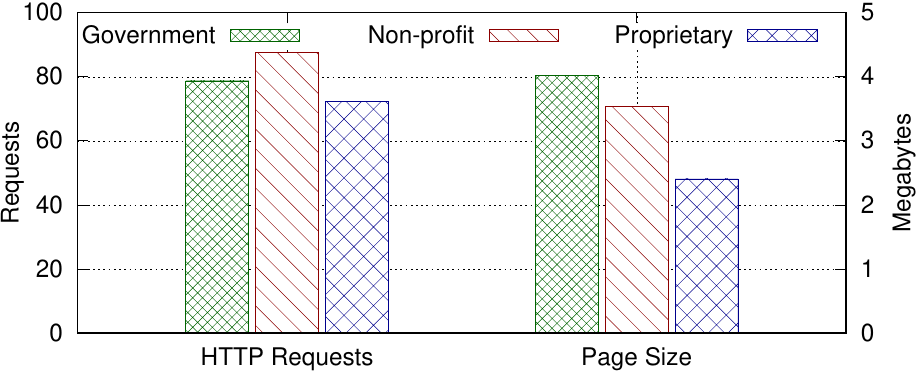}
    \caption{Request and response size comparison.}
    \label{fig:page_size_http}
\end{figure}

\BfPara{HTTP Request}
The HTTP (Hypertext Transfer Protocol) request is a message that is sent from a client (such as a web browser) to a server, requesting a particular resource or action. The request typically includes a URL (Uniform Resource Locator) that specifies the resource or action being requested, along with any additional information needed by the server to fulfill the request, such as headers and cookies.

By Analyzing HTTP requests, businesses and organizations can gain insights into the performance of their websites, including, for example, the speed of page load times, the number of requests per page, and the size of files being requested. This information can help identify areas where website performance can be further optimized.

Figure~\ref{fig:page_size_http} shows the average HTTP requests per website across the three different categories of analyzed hospitals. We found that most websites generated 65 to 90 HTTP requests per visit, with the non-profit hospitals being the highest. Despite having relatively similar HTTP requests, the proprietary hospitals' average page size was 45\% less than the government hospitals. Upon further analysis, we found that the proprietary hospitals' websites contain the least percentage of images in contrast to the government and non-profit hospitals, which help explain this trend.

%\noindent\textit{\textbf{\underline{Key Takeaway:}}} 
\takeaway{({\bf RQ1.}) Structurally, the content type distribution of the hospitals' categories are similar, except for \textit{``Image''} and \textit{``Script''} content types. Although the HTTP requests were relatively similar among hospitals, the average page size of the proprietary hospitals was 45\% smaller than that of the government hospital.} 
%% add the takeaway about http? contrast to anything out there. 

\begin{figure}[t]
  \centering
  \includegraphics[width=0.9\linewidth]{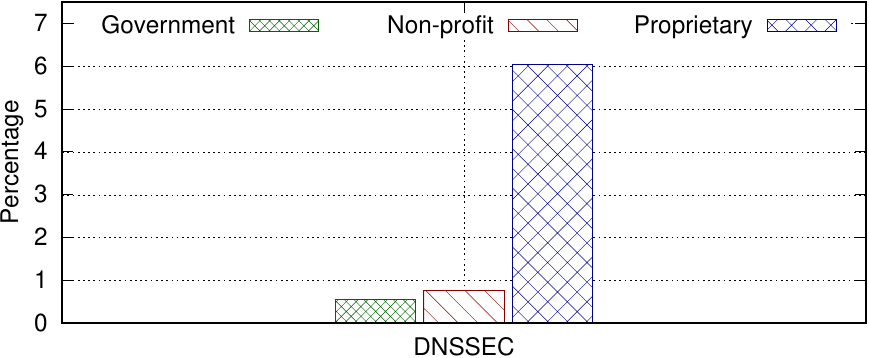}
  \caption{Domain Name System Security Extensions.}
  \label{fig:dnssec}%
\end{figure}

\begin{table}[t]
\centering
\caption{The corresponding certificate issuer organizations for the hospitals' websites. Notice that \textit{Let's Encrypt} is the most prominent certificate issuer organization, with up to 25.21\% associated URLs.}
\label{Tab:certificate_issuer}
\begin{adjustbox}{width=\columnwidth,center}
\begin{tabular}[t]{lrrrrrrr}
\Xhline{2\arrayrulewidth}
\multirow{2}{*}{Issuer Organization} & \multirow{2}{*}{Paid} & \multicolumn{2}{c}{ Government}   & \multicolumn{2}{c}{Non-profit} & \multicolumn{2}{c}{proprietary} \\
\cline{3-8}
 & & \multicolumn{1}{c}{ \#} & \multicolumn{1}{c}{\%} & \multicolumn{1}{c}{ \#} & \multicolumn{1}{c}{\%} & \multicolumn{1}{c}{\#} & \multicolumn{1}{c}{\%} \\
\Xhline{2\arrayrulewidth}
Let's Encrypt & \xmark & 321   &  31.04  & 559   &  25.56  & 295    &   19.03  \\
GoDaddy.com\_Inc. & \checkmark  & 131  &  12.66  & 175  &  8.00 & 132   &   8.51 \\
Cloudflare\_Inc. & \xmark & 25  &  2.41  & 108  &  4.93 & 179 &  11.54 \\
Sectigo Limited & \checkmark & 67 & 6.47 & 133  & 6.08 & 38 &  2.45\\
cPanel\_Inc. & \xmark & 47  & 4.54 & 72  & 3.29 & 111 & 7.16 \\
DigiCert Inc.  & \checkmark & 35 &  3.38 & 154 & 7.04  & 14 &  0.90\\
Trustwave Holdings, Inc.   & \checkmark & 1 & 0.09 & 9 & 0.41 & 148 &  9.54\\
Entrust L1K & \checkmark & 9   & 0.87 & 86 & 3.93 & 43 &  2.77 \\
Other & - & 148 & 14.31 & 385 & 17.60 & 149 &  9.61\\
\Xhline{2\arrayrulewidth}
No SSL Certificate Found & - & 250 &  24.17 & 506 & 23.13 & 441 &  28.45  \\
\Xhline{2\arrayrulewidth}
\end{tabular}
\end{adjustbox}
\end{table}

\subsection{DNSSEC Prevalence}\label{sec:dnssec} DNSSEC (Domain Name System Security Extensions) is a security protocol that is used to protect against certain types of attacks on the Domain Name System (DNS), targeting the integrity and source authenticity.

DNSSEC works by adding digital signatures to the DNS records, which DNS resolvers can verify to ensure that the records have not been tampered with or forged. This helps to prevent attackers from redirecting users to fake websites, intercepting email messages, or other types of attacks that rely on manipulating DNS records. DNSSEC is designed to provide end-to-end security, meaning that the integrity of the DNS records can be verified from the root servers down to the individual domain name servers. It is supported by most modern web browsers and operating systems, and many top-level domains have already adopted DNSSEC to provide an additional layer of security for their users.

To gauge its prevalence in this rather critical category of websites, we investigated the DNSSEC deployment and found that 6\% of the proprietary hospital websites enabled DNSSEC as shown in Figure~\ref{fig:dnssec}, which is significantly higher than government (0.55\%) and non-profit (0.75\%) hospital websites.

\takeaway{({\bf RQ2.}) We notice that 6\% of proprietary hospitals used DNSSEC compared to under 1\% in government and non-profit. While 1\% is considered a small percentage, it is consistent with the DNSSEC deployment in {\tt .com} and {\tt .org} general websites (i.e., 0.75\%--1\%)~\cite{chung2017longitudinal}, whereas the proprietary group has 6 to 8 times more than those levels, highlight better prevalence in this category.}

\subsection{HTTPS and SSL Certificate Analysis}\label{sec:sslcetrificate}
The HTTP protocol is responsible for transferring website content from the web server to the endpoint browser. However, this protocol is insecure, exposing content to unauthorized access. Therefore, most websites have moved to using HTTPS, a secure version of HTTP, on top of the Secure Sockets Layer (SSL), which, among other functions, implements an encryption mechanism to protect the transferred content between web servers and endpoint browsers. Healthcare websites often require users to enter sensitive personal information such as health-related data, insurance information, and medical history. HTTPS can help protect this information from being intercepted by attackers, thereby ensuring that patient data remains confidential.

To this end, we next look into SSL-related configurations: certificate authority, signature algorithm, and certificate validation. Among the studied hospitals, we noticed that 25.25\% of the websites are still using HTTP, in contrast to only around 20\% in general web~\cite{W3Techs}. While there is a 5\% of difference in the number of websites, that number is alarmingly high given the type of data associated with hospitals (i.e., at least one out of four hospitals uses an insecure protocol).

\BfPara{Certificate Authority}
A certificate authority (CA) is an organization that validates the identities of entities, including websites, email addresses, etc., by binding entities to cryptographic keys through the issuance of electronic documents. 
Investigating the certificate authority organization (i.e., the issuer of the certificate), Table~\ref{Tab:certificate_issuer} shows that the majority of hospitals are using the free \textit{Let's Encrypt} services~\cite{lets} (i.e., free SSL certificates)., with up to 31.04\% for the governmental hospital group. We also notice that hospitals' websites widely use free SSL certificates. Surprisingly, we did not find SSL certificates in 24.17\% of government, 23.13\% of non-profit, and 28.45\% of proprietary websites.

% \textcolor{blue}{It is noteworthy to mention that SSL certificates were absent in a significant percentage of websites belonging to different sectors. Specifically, 24.17\% of government, 23.13\% of non-profit, and 28.45\% of proprietary websites lacked SSL certificates. This lack of SSL certificates can potentially compromise the websites' security and their users' sensitive information.}

\begin{table}[t]
\centering
\caption{SSL signature algorithms' comparison.}
\label{Tab:signature algorithms}
\begin{adjustbox}{width=\columnwidth,center}
\begin{tabular}[t]{lrrrrrr}
\Xhline{2\arrayrulewidth}
 & \multicolumn{2}{c}{Government} & \multicolumn{2}{c}{Non-profit} & \multicolumn{2}{c}{proprietary} \\
%  & Government && Non-profit && proprietary \\
\Xhline{2\arrayrulewidth}
Algorithms & \# & \% &  \# & \% &  \# & \%  \\
\hline
SHA256 with RSA &751&95.00     &1,535&91.31      &891&80.34\\
SHA256 with ECDSA &26&3.30   &108&6.42       &179&16.14\\
SHA384 with ECDSA &10&1.27    &35&2.08            &37&3.34\\
SHA1 with RSA &1&0.13         &2&0.12          &1&0.09\\
SHA384 with RSA &1&0.13       &-&-          &1&0.09\\
SHA512 with RSA &-&-          &1&0.06          &-&-\\
\Xhline{2\arrayrulewidth}
\end{tabular}
\end{adjustbox}
\end{table}%

\BfPara{Algorithms} Table~\ref{Tab:signature algorithms} shows the SSL signature algorithms used by the government, proprietary, and non-profit websites. As shown, \textit{SHA256 with RSA} is the most used scheme with 95.00\% for government, 91.31\% for non-profit, and 80.34\% for proprietary hospital websites, respectively. This is mainly because hospitals intend to use traditional go-to algorithms adopted by service providers. On the other hand, we notice that fewer hospitals website use \textit{SHA256 with ECDSA} (Elliptic Curve Digital Signature Algorithm) algorithm that uses shorter keys for the same security level as in RSA with larger keys \cite{alabduljabbar2022certificates}. With 3.30\% for government, 6.42\% for non-profit, and 16.14\% for proprietary hospital websites, respectively. We note ECDSA is a newer and more efficient algorithm and is mainly used in newer websites \cite{alabduljabbar2022certificates}. ECDSA is, however, more vulnerable to attacks than the older RSA under post-quantum adversaries, according to recent studies \cite{RoettelerNSL17}.

\begin{figure}[t]
  \centering
  \includegraphics[width=0.99\linewidth]{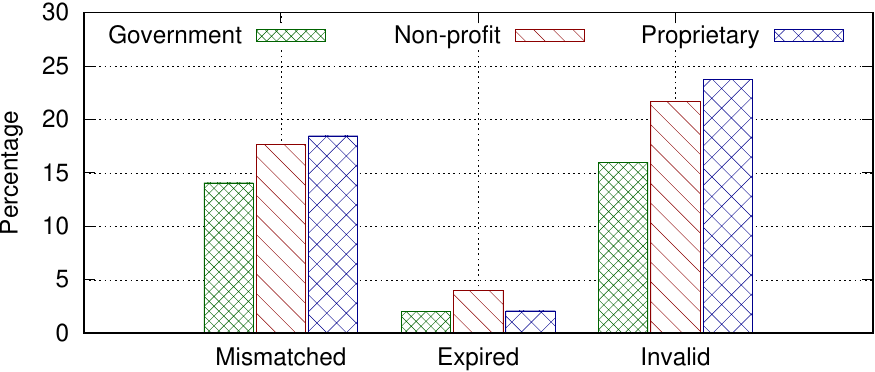}
  \caption{The SSL validity comparison of Government, Non-profit, and proprietary hospital websites.}
  \label{figure:SSL_Valid}
\end{figure}

\BfPara{Certificate Validity}
We further investigated the SSL certificate validity and potential issues. In the following, we discuss issues related to SSL certificate failures (see Figure~\ref{figure:SSL_Valid}).

\noindent\textit{SSL Mismatched Domain.} A mismatched domain might be an indication of website impersonation or inconsistent website migration, and both highlight a lack of rigorous security practices. We found that 18.45\% of the proprietary hospitals had SSL certificates with mismatched domains, versus 14.05\% of the government hospitals and 17.67\% of the non-profit hospitals. Even varying, all hospitals' websites had concerning ratios of mismatched domains. 

\noindent\textit{SSL Expired.} Our analysis uncovered that about 3.97\% of non-profit hospitals have expired certificates, compared to 2.20\% and 2.09\% for government and proprietary hospitals. Similar to our previous analysis of out-of-date websites, this may lead to potential user information and data privacy risks.

\noindent\textit{SSL Invalid.} The invalidity of SSL means that some fields in the certificate are incorrect. Surprisingly, all hospital categories had an alarming percentage of invalid SSL certificates, with 15.94\% for government, 21.70\% for non-profit, and 23.72\% for proprietary hospital websites.

%\noindent\textit{\textbf{\underline{Key Takeaway:}}} 

\takeaway{({\bf RQ2.}) More than 25\% of hospitals' websites are using the plain HTTP protocol, which is alarmingly higher than $\approx$20\% in the general websites~\cite{W3Techs}. 
Among websites that used HTTPS, 88.77\% of them used \textit{SHA256 with RSA}. 
Among the $\approx$75\% hospitals with an SSL certificate, we found that 20.45\% of the SSL certificates were invalid while 16.72\% had a mismatched domain name, primarily in proprietary hospitals in both cases.}

  \begin{figure}[t]
  \centering
  \includegraphics[width=0.99\linewidth]{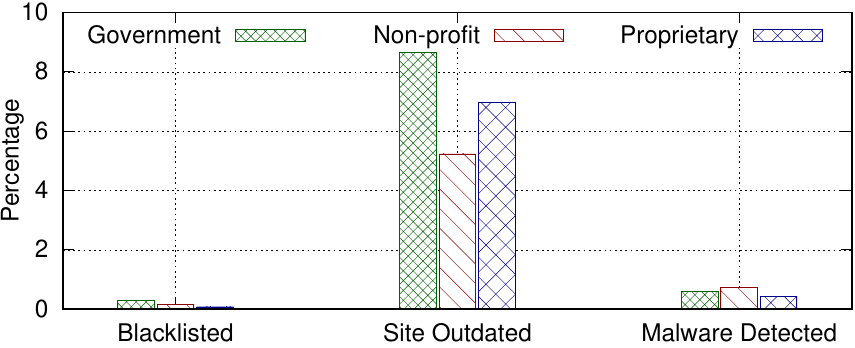}
  \caption{Comparing the maliciousness of Government, Non-profit, and proprietary hospitals' websites.}
  \label{figure:sucuri}
\end{figure}

\subsection{Malicious Activities Analysis}\label{sec:malicious}
In addition to the structural differences and SSL certificate analyses, we study the malicious activities associated with the hospitals' websites. Malicious activities considered in this work include providing malicious or phishing content or the association of website resources with malicious attacks.

\BfPara{Domain-based Malicious Activities}
We leveraged Sucuri \cite{Sucuri} to explore domain-based malicious activities. Figure~\ref{figure:sucuri} shows that although only a small portion of hospitals' URLs are blacklisted or labeled as malware, about 8.66\% of government, 5.21\% of non-profit, and 6.96\% of proprietary hospital websites are outdated, which raises concerns of data leakage.

\begin{figure}[t]
  \centering
  \includegraphics[width=0.99\linewidth]{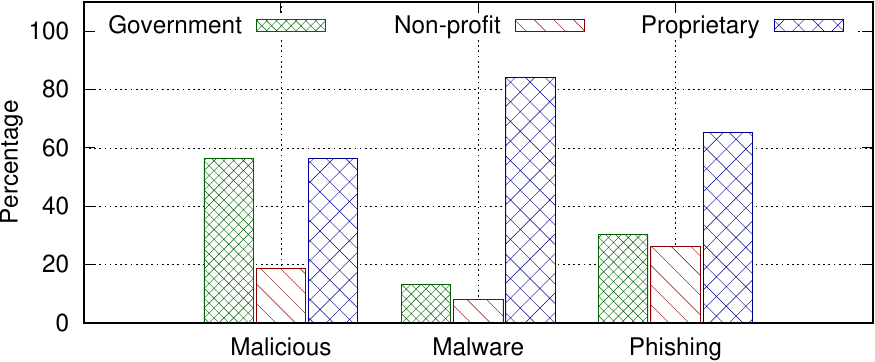}
  \caption{The potential maliciousness of Government, Non-profit, and proprietary hospitals.}
  \label{figure:VirusTotal_Labels}
\end{figure}

\BfPara{Content-based Malicious Activities}
Next, we analyzed the website content using VirusTotal API~\cite{VirusTotal}. Figure~\ref{figure:VirusTotal_Labels} shows that, among the three hospital categories, 84.21\% of proprietary hospitals contained malware, compared to only 13.15\% and 7.89\% in the government and non-profit hospitals, respectively. Moreover, we observed that 65.21\% of proprietary hospital websites are suspected of having phishing-like behaviors. We note that the percentage for the government (30.43\%) and non-profit (26.08\%) hospitals are significantly smaller than that of the proprietary hospitals but still noticeably high. Besides, we observed that 56.25\% of government and proprietary hospitals and 18.75\% of non-profit hospitals are associated with malicious activities.

%\noindent\textit{\textbf{\underline{Key Takeaway:}}} 
\takeaway{({\bf RQ2.}) Most hospitals have maliciousness features (domain or content), and many are vulnerable to data leakage due to a lack of maintenance (i.e., outdated websites). 
Among the compiled hospital websites, 8.66\% of the websites are outdated. In addition, a concerning portion of websites contains malicious content and is associated with phishing and malicious behaviors.}

\section{Data Breaches Analysis}\label{sec:databreach}
Analyzing the data breaches helps understand the correlation between web presence security and incidents. According to the Health Insurance Portability and Accountability Act (HIPPA), a data breach can be defined as an impermissible use or disclosure under the Privacy Rule that compromises the security or privacy of the protected health information \cite{HIPPA}. In the healthcare domain, data breaches are devastating, as they cause damage to patients and healthcare organizations alike. Recent works have shown that healthcare is the most targeted industry by cyber criminals due to financial gain as attackers intend to sell patients' records on the dark web. 
To investigate historical data breach incidents in hospitals, we obtained the healthcare data breaches dataset from the U.S. Department of Health and Human Services, Office for Civil Rights (OCR). The OCR portal lists all data breaches of unsecured health information affecting 500 or more patients. The OCR portal categorizes data breaches into two categories; (i) incidents reported within the last 24 months and currently under investigation and (ii) the achieved breaches, which comprise the resolved breach reports older than 24 months. We note that it is challenging to associate the hospital names with the entities named in the data breaches, as they are not consistently organized (e.g., mixture or truncation). 
To resolve the issue, we started by using the hospitals' names as anchors and then leveraged Natural Language Toolkit (NLTK) \cite{NLTK} for punctuation removal, case normalization, stopwords removal, and lemmatization \& stemming of the hospitals' names. A similar process was followed for the entity name among the data breaches dataset.
Lastly, any hospital name and entity with two common words are filtered for manual analysis and vetting. Overall, we manually inspected 1,253 incidents, resulting in 414 accurate labeling of data breaches.

Hereafter, we analyze the data breaches, providing insights into the common online attributes of the breached hospitals.

\BfPara{Associated Hospitals \& Individuals}
Among the 414 data breach incidents in our dataset, 49 were government hospital-related, 156 were non-profit hospital-related, and 34 were proprietary hospital-related. It is worth mentioning that a hospital may be involved in several incidents. Our analysis indicates that the average number of affected individuals is 58,750 overall, including 60,458 for government, 64,977 for non-profit, and 50,815 for proprietary hospitals. Remarkably, the proprietary hospitals are involved in the least number of incidents and affected individuals.

\begin{figure}[t]
  \centering
  \includegraphics[width=0.99\linewidth]{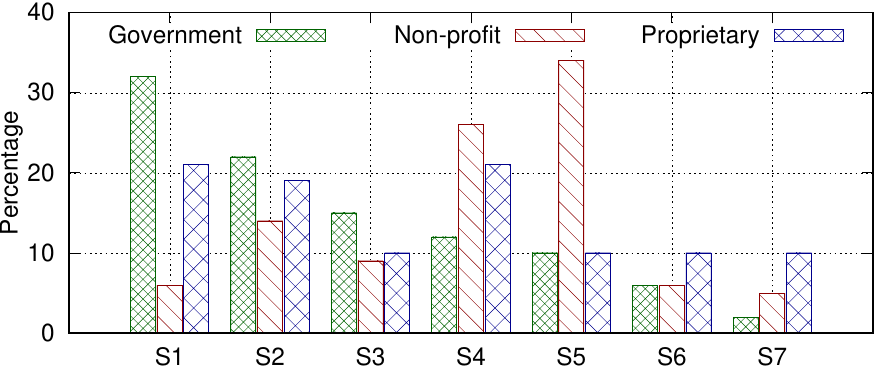}
  \caption{Comparing the data breach surfaces of Government, Non-profit, and proprietary hospitals. S1= Paper/Films, S2=Network Server, S3=EMR, S4=Other, S5=Email, S6=Laptop, S7=Desktop Computer.}
  \label{fig:Breached_information}
\end{figure}

\BfPara{Data Breach Surface} As shown in Figure~\ref{fig:Breached_information}, \textit{``paper/films''} are the most commonly targeted for government (32\%) and proprietary (21\%) hospitals, despite only 6\% for non-profit hospitals. Then, \textit{``emails''} are mostly targeted in non-profit hospitals (34\%), despite not being heavily targeted in government and proprietary hospitals (1\%).
Overall, \textit{``network server''} is the second most common target after \textit{``paper/films''}, inferring the importance of hospitals' online security.

  \begin{figure}[t]
  \centering
  \includegraphics[width=0.99\linewidth]{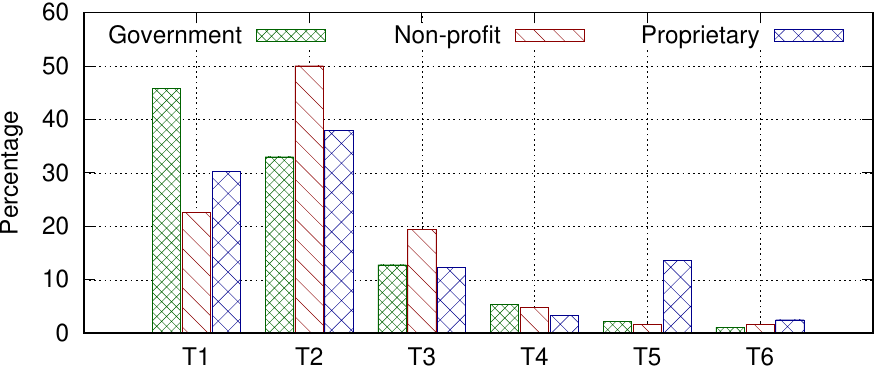}
  \caption{Comparing the data breach types of Government, Non-profit, and proprietary hospitals. T1=Hacking/IT, T2=Unauthorized Access, T3=Theft, T4=Loss, T5=Improper Disposal, T6=Other.}
  \label{figure:incidents_type}
\end{figure}
%TODO

\BfPara{Data Breach Type} As shown in Figure~\ref{figure:incidents_type}, we found the majority of incidents are \textit{``hacking/IT''} representing 45.75\% in the government hospitals, followed by proprietary and non-profit hospitals representing 30.22\% and 22.58\%, respectively. Further, we observed that \textit{``unauthorized access/disclosure''} is the most common data breach type within the non-profit hospitals representing 50\%, while 37.91\% for proprietary and 32.98\% for government hospitals.

\begin{table}[t]
\caption{Attributes extracted for data breach analysis.}
\label{tab:features_description}
\centering
\scalebox{0.81}{
\begin{tabular}{lll}
\Xhline{2\arrayrulewidth}
 & Title & Description \\
\Xhline{2\arrayrulewidth}
F1 & Certificate Invalid & The browser fails to verify website certificate   \\
F2 & Certificate Unmatched & The website name does not match SSL certificate  \\ 
F3 & Certificate\_Expired & The website certificate becomes invalid  \\
F4 & Validity\_Days\_Left & The remaining validity days of website certificate \\
F5 & Positives & The website domain is detected by VirusTotal API   \\
F6 & Malicious\_Site &  Websites detected as malicious by VirusTotal API\\
F7 & Malware\_Site &  Websites detected as by malware VirusTotal API\\
F8 & Phishing\_Site & Websites detected as  by VirusTotal API phishing \\
F9 & Page\_Size (MB) & The website average page size of in MegaByte \\
F10 & Load\_Time (S) & The website average page load time of in seconds \\
F11 & Number of Requests & The website average number of requests \\
F12 & CSS  & The percentage of CSS retrieved by Pingdom API \\
F13 & Font & The percentage of font retrieved by Pingdom API\\
F14 & HTML & The percentage of HTML retrieved by Pingdom API\\
F15 & Image & The percentage of images retrieved by Pingdom API \\
F16 & Redirect & The percentage of redirect retrieved by Pingdom API\\
F17 & Script & The percentage of script retrieved by Pingdom API\\
F18 & XHR & The percentage of XHR retrieved by Pingdom API \\
F19 & Blacklisted Flag & Websites detected as blacklisted by Sucuri API \\
F20 & Malware Flag & Websites detected as malware by Sucuri API \\
F21 & DNSSec Flag & Websites detected using DNSSec \\
\Xhline{2\arrayrulewidth}
\end{tabular}}
\end{table}

\BfPara{Data Breach Online Presence Attribution}
To understand the relationship between online presence, security properties, and data breach incidents, we used a gradient boosting model with non-negativity constraint (i.e., monotonously constraint) to learn important attributes of breach incidents.  

\begin{figure}[t]
  \centering
  \includegraphics[width=0.99\linewidth]{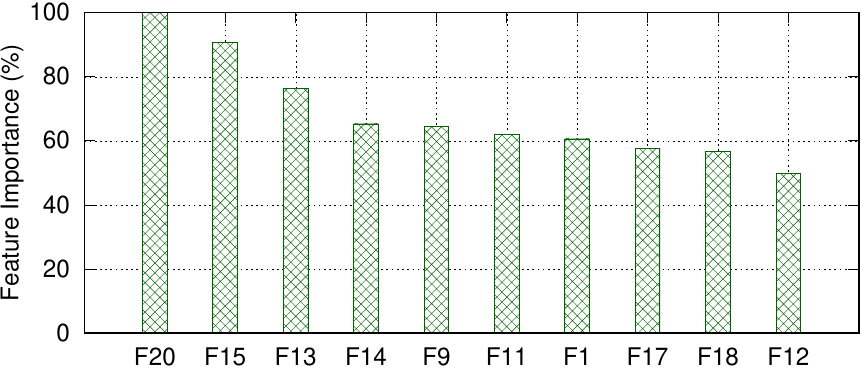}
  \caption{The domain- and content-level attributes importance (\%) in distinguishing hospital websites associated with data breaches. The titles and descriptions of all features are shown in Table~\ref{tab:features_description}}
  \label{figure:importance_Features}%
\end{figure}

Table~\ref{tab:features_description} illustrates the 21 attributes used in our model, and Figure~\ref{figure:importance_Features} shows the ten attributes that are directly (and mostly) correlated with the breached websites. As shown in the figure, when the website contains malware software (F20: Websites detected as malware by Sucuri API), it is (naturally) more likely to be involved in a data breach incident. Other features that highly correlated with websites' data breaches are (F15: The percentage of images retrieved by Pingdom API, and F13: The percentage of font retrieved by Pingdom API).

%\noindent\textit{\textbf{\underline{Key Takeaway:}}} 
\takeaway{({\bf RQ2.}) 156 non-profit hospitals were associated with reported breach incidents, which is significantly higher than the remaining two categories. Moreover, hospitals' online presence security features play a clear role in their potential to be targeted, according to the top 10 attributes of hospital websites indicative of data breach incidents. We also notice that proprietary hospitals are the least susceptible to breaches. 
}

\section{Concluding Remarks}
Recent reports showed an increasing trend of attacks targeting hospital networks to compromise sensitive patient data. 
In this paper, we investigated the online presence of hospitals by analyzing their websites. Benefiting from a categorization into government, non-profit, and proprietary hospitals, we conduct a comparative study that sheds light on various structural and security features. 
Of particular note, we investigated the SSL certificate validity, the related issues among hospitals' websites, and malicious associated behaviors.  
Leveraging the collected attributes as features, we demonstrate the most important attributes indicative of websites associated with data breach incidents and helpful in understanding their security.
Our findings are among the first steps toward achieving patient security, alarmingly highlighting the lax security in many hospitals' websites.

%\bibliography{ref.bib}
%\bibliographystyle{plain}

\end{document}